\documentstyle[aps,prl]{revtex}

\author{Andr\'es Jord\'an
 and Mat\'{\i}as Libedinsky}

\title{Path Integral Invariance under Point Transformations}

\address{Departamento de F\'{\i}sica, Facultad de Ciencias, Universidad
de Chile, Casilla 653, Santiago 9, Chile\\
and\\
Centro de Estudios Cient\'{\i}ficos de Santiago, Casilla 16443, Santiago,
 Chile}

\begin{document}

\draft

\maketitle

\begin{abstract}

We give here a covariant definition of the path integral formalism for
the Lagrangian, which leaves a freedom to choose anyone of many possible
quantum systems that correspond to the same classical limit without
adding new potential terms nor searching for a strange measure, but
using as a framework the geometry of the spaces considered. We focus our
attention on the set of paths used to join succesive points in the
discretization if the time-slicing definition is used to calculate the
integral.If this set of paths is not preserved when performing a point
transformation, the integral may change. The reasons for this are
geometrically explained. Explicit calculation of the Kernel in polar
coordinates is made, yielding the same system as in Cartesian coordinates.

\end{abstract}

\pacs{03.65.Ca}

It has been argued that point  transformations performed by simply
changing integration variables in the path integral lead to inequivalent
results. For example Edwards and Gulyaev \cite{EG} had shown
 that the free particle path integral in Cartesian coordinates is not
equal to the naive expresion in polar coordinates.

 This result is also quoted by several authors \cite{S,GS,GJ,OC,MS,CCG},
who argued that new terms should be added to the  action in the new
coordinates beyond what would be expected clasically. The same is
pointed out in the book of R.J. Rivers \cite{R} , tracing the problem to
the stochastic nature of quantum paths.

 The difference between the results was explained by assuming that
different orderings were given to the Hamiltonian operator in each case.
 This result suggests that the quantum theory inevitably depends on the
choice of coordinates. While there is no guarantee that quantum mechanics
should respect a classical feature such as general covariance, it is hard
to believe that a physical result should depend on the coordinates used to
 parametrize the states. Following this philosophy, DeWitt \cite{DeW}
developed a covariant quantization method showing that there exists a
way of changing coordinates without the system being changed. What we do
here is to develop a covariant definition of the path integral that
leaves a freedom to choose one from many possible quantum system that
correspond to the same classical limit.

Consider the following definition for the  path integral in Cartesian
coordinates in two dimensions \cite{FH,F} 

\begin{equation}
 K(b,a)=\lim_{\epsilon \rightarrow 0} \frac{1}{A^{2n}}\int \cdots \int
e^{\frac{i}{\hbar} S[b,a]}\,dx_{1}dy_{1}\ldots dx_{n-1}dy_{n-1},
\label{eq:1}
\end{equation}
where

\begin{equation}
S[b,a]=\sum^{n}_{k=1}S_{cl}[x_{k},y_{k},t_{k};x_{k-1},y_{k-1},t_{k-1}].
\label{eq:2}
\end{equation}

In this last expression, $S_{cl}$ is the classical action evaluated for
a path connecting $x_{k-1},y_{k-1}$ and $x_{k},y_{k}$ (micropath), and
$[b,a]$  represent the end points $(\vec{x}_b,t_{b};\vec{x}_a,t_{a})$ .
The time interval $T=t_{b}-t_{a}$ is divided into $n$ intervals  of
length $\epsilon=T/n$, the integration is  over all the possible
positions taken at each time and $A=\left(\frac{2\pi i\hbar
\epsilon}{m}\right)^{1/2}$ is the normalizing constant. Finally, the
limit is taken.

The assignment $\phi [x(t)]=e^{\frac{i}{\hbar}\int
{\cal{L}}(\dot{x},x,t)\,dt}= e^{\frac{i}{\hbar}S[x(t)]}$ might lead one
to think that only the Lagrangian is needed to carry out the calculation
of the path integral in Cartesian coordinates. However, this is not true
because for non differentiable paths $\dot{x}$ doesn't exist and thus
the Lagrangian is ill-defined. These paths are important because they
give the largest contribution to the path integral \cite{F}. Therefore,
for these non-differentiable paths the number assigned to them  may be
calculated in the following way:

\begin{equation}
 S[x(t)]=\lim_{\epsilon \rightarrow 0}\sum_{k=1}^{n}S_{cl}[x(t_{a}+k\epsilon),
x(t_{a}+(k-1)\epsilon],
\label{eq:3}
\end{equation}
where
\begin{equation}
S_{cl}[x(t_{a}+k\epsilon),x(t_{a}+(k-1)\epsilon)]=
\int_{t_{a}+(k-1)\epsilon}^{t_{a}+k\epsilon}
{\cal{L}}(\dot{\psi}_{k},\psi_{k},t)\,dt ,
\end{equation}
 here $\psi_{k}(t)$ describes the micropath selected to join
$x(t_{a}+(k-1)\epsilon)$ to $x(t_{a}+k\epsilon)$. 

We can see here that for defining the path integral one needs not only the
Lagrangian but also a set of paths to join points of the discretized trajectory.
This gives one the freedom of choosing different ones, which may lead to
inequivalent quantum systems. 

For example, consider the Lagrangian  in one dimension,
${\cal{L}}=\sqrt{1+\dot{x}^{2}}$, which describes a free particle, and
the path $x(t)\equiv 0$ from $t=0$ to $t=l$.We want to know which value
will be asigned to this path. If one chooses for the $\psi_{k}$ straight
lines and calculate $S[x(t)]$ by limiting procedure (\ref{eq:3}) one 
finds $\phi [x(t)]=e^{\frac{i}{\hbar}l}$. But, if for the $\psi_{k}$ one
chooses semicircles and you do limiting procedure (\ref{eq:3}) one finds
$\phi [x(t)]=e^{\frac{i}{\hbar}\pi l/2}$ (see FIG.\ref{fig1}).

 Now, in this example we picked up the simplest of all paths, a straight
line, and with two different limiting procedures, we found two different
values. The reason for this is that with straight lines, semicircles or
any other kind of paths, we can approximate pointwise to the original
path, but not necesarilly the derivatives have to converge to the
derivatives of the original path. So, on a diferentiable path, one could
say that a ``good" way to calculate $S[x(t)]$ is to choose paths
$\psi_{k}$ that in the limit will approximate not only pointwise but
also on the derivatives. But a bigger problem arises on the
non-differentiable paths, because there we can't define a ``good" way,
and the value given to the path $x(t)$ depends exclusively on which
paths $\psi_{k}$ we select. So here one can clearly see that care must
be taken about which micropath one chooses. In \cite{FH},pg. 34, it says
``It is possible to define the path in a somewhat more elegant manner.
Instead of straight lines between the points $i$ and $i+1$, we could use
sections of the classical orbit. ". But we can see in our example that
changing the micropaths may change the final result, if it is not done with
 care. And although Feynmann introduces this to win in elegance, one can
reintroduce elegance simply by defining the ``straight line" as a
geodesic.  
 
Now, when a change of variables is made, one must preserve the paths
$\psi_{k}$ used to join the points in the discretization, because if one
doesn't do that the values assigned to a specific path will change and the
final result will differ.  The paths are geometric objects independent of
 the coordinates and this is the basis of our covariant aproach. This
includes choosing geodesics in the configuration space, or the classical
paths, as the $\psi_{k}$. Later we show how to calculate in
polar coordinates the kernel of the free particle, giving the same
result as in Cartesian, with the condition that one doesn't change the
values given to the paths or, what is the same, the $\psi_{k}$ selected
are the same.

There is another source of confusion when evaluating a path integral.
Consider for instance the free particle in Cartesian coordinates; it can
be shown \cite{MaDo} that if the Weyl ordering is considered the kernel of
the Schrodinger equation can be written as:

\begin{equation}
K(b,a)=\lim_{\epsilon \rightarrow 0}\frac{1}{A^{2n}}\int \cdots
\int e^{\frac{i}{\hbar}\sum_{k=1}^{n}\left[(x_{k}-x_{k-1})^{2}/\epsilon+
(x_{k}-x_{k-1})^{2}/\epsilon - \epsilon V((x_{k}+x_{k-1})/2,(y_{k}+y_{k-1})/2)
\right]}dx_{1}dy_{1}\ldots dx_{n-1}dy_{n-1}.
\label{eq:5}
\end{equation}

 This expresions, however, may be wrong if we consider another ordering
or curvilinear coordinates. This expression is called the mid-point
definition. There are other common definitions known in literature as
left or right-point which are studied by Feynman \cite{F}. This
definitions are not equivalent but they may yield the same result if the
conditions discussed below are fulfilled.

In general $S_{cl}[x_{k},y_{k};x_{k-1},y_{k-1}]$ can be approximated so long as
the error  is of order $\epsilon^{1+\alpha}$ with $\alpha
>0$, i.e., if a function
$\bar{S}[x_{k},y_{k};x_{k-1},y_{k-1}]$ can be defined such that

\begin{equation}
S_{cl}[x_{k},y_{k};x_{k-1},y_{k-1}]=\bar{S}[x_{k},y_{k};x_{k-1},y_{k-1}]
+O(\epsilon^{1+\alpha}) \label{eq:6} \end{equation} 
calculated as (\ref{eq:1}) replacing $S_{cl}$ by $\bar{S}$ in
(\ref{eq:2}). With this result, and starting from the path integral
formalism (without mentioning an order in the Hamiltonian operator),
result (\ref{eq:5}) may be also understood: if we choose as micropaths
the straight lines, the integral of the Lagrangian through this
micropaths can be approximated by the expression on the exponent in
(\ref{eq:5}) with an error $O(\epsilon^{2})$ by the paralelogram law. So
we can associate the Weyl ordering in Cartesian coordinates with this
selection of micropaths.  

  So, talking about the mid-point definition  (\ref{eq:5}) of path
integrals can be misleading. The integral doesn't change  if one chooses
the right, or left, or midpoint definition while condition (\ref{eq:6})
is satisfied; they are only a tool for calculation. Any of these may be
used as defining the path integral, because they provide one with a
function that gives a number when you have two points in the discretized
path, and so one can proceed in the calculation,  but they will not
agree with (\ref{eq:5}) if condition (\ref{eq:6}) is not satisfied.

Consider the kernel of the free particle in polar coordinates.
The explicit expresion  reads

\begin{equation}
K(b,a)=\frac{1}{A^2} \lim_{\epsilon \rightarrow
0}\int^{\infty}_{0}r_{1}\,\frac{dr_{1}}{A}\cdots \int^{\infty}_{0}
r_{n-1}\,\frac{dr_{n-1}}{A}\int^{2\pi}_{0}\frac{d\theta_{1}}{A}\cdots
\int^{2\pi}_{0}\frac{d\theta_{n-1}}{A}\exp
\left[\frac{i}{\hbar}S(b,a)\right].
\label{eq:ob}
\end{equation}

 The action can
be decomposed into the sum of the actions for each segment of path.
For every one of such pieces the action must be evaluated on a prescribed
trajectory. We choose these trajectories to be the classical ones
(straight lines) and in polar coordinates this gives the action for the
total path as 

\begin{equation} 
\exp \left[\frac{i}{\hbar}S_{cl}(b,a)\right]=\prod_{k=1}^{n-1}\exp
\left[ \frac{im}{2\hbar
\epsilon}(r_{k}^{2}+r_{k-1}^2-2r_{k}r_{k-1}\cos(\theta_{k}-\theta_{k-1})
\right]. 
\end{equation}

Note that the micropaths choosen are the geodesics in the polar plane,
that in this case coincides with the classic trajectory of the particle;
we don't need to know what was the result of quantization in Cartesian:
we only need the Lagrangian and the geometry of the new coordinates. As
we go over the paths, $\theta_{k}-\theta_{k-1}$ need not to be small.
This difference can take all values between $0$ and $2\pi$, so one can't
expand the cosine and keep only the lowest order terms as is done in
\cite{EG}. In order to calculate the kernel we first collect the terms
in an integral  of the general form $\int^{2\pi}_{0}\exp\left[i(C\cos
\theta+ S\sin \theta)\right]\,d\theta$ which can be readily done by
integrating over the unit circle in the complex plane and calculating
the residues. The result is $\frac{2\pi}{A}J_{0}(\sqrt{C^2+S^2}) $,
where $J_{0}$ is the Bessel function of the first kind of order $0$.
 Next we collect the terms in $r_{1}$ obtaining
an integral that can be worked out using the following result \cite{GR}

\begin{equation} 
\int^{\infty}_{0}x^{\nu +1}e^{\pm i\alpha x^{2}}J_{\nu}(\beta
x)\,dx=\frac{\beta^{\nu}}{(2\alpha)^{\nu +1}}\exp\left[\pm i\left(\frac{\nu
+1}{2}\pi - \frac{\beta^{2}}{4\alpha}\right)\right] 
\end{equation}

where $ \alpha \, , \beta >0$  and $-1<Re(\nu)<1/2$. Using this formula
with $\nu=0$, $\alpha=m/\hbar \epsilon$ and $\beta=(m/\hbar \epsilon)
\sqrt{r_{0}^{2}+r_{2}^{2}+2r_{0}r_{2} \cos(\theta_{2}-\theta_{0})}$ ,
gathering all the terms in $r_{0}\, ,r_{2}$ and replacing the apropriate
constants we obtain  the integrals over $r_{1}$ and $\theta_{1}$ and
continuing with this process after $n-1$ integrations we obtain

\begin{equation} 
\frac{1}{n}\exp\left(\frac{im}{n\hbar \epsilon}\left(\frac{n+1 }{2}r_{n}^{2}
+\frac{1}{2}r_{0}^{2}-r_{n}r_{0}\cos(\theta_{n}-\theta_{0})\right)\right).
\end{equation}

We have to multiply now by the term $\exp(-imr_{n}^{2}/2\hbar
\epsilon)$ , because we were multipliyng each time by
$\exp(imr_{k}^{2}/\hbar \epsilon)$, but because $r_{n}$ is an end point we
 have only $\exp(imr_{n}^{2}/2\hbar \epsilon)$. So, using the fact that
$n\epsilon=T=t_{b}-t_{a}$ we finally obtain for the kernel in polar
coordinates:

\begin{equation}
K(b,a)= \frac{m}{2\pi \hbar Ti}\exp\left[\frac{im}{2T \hbar}
\left(r_{n}^{2}+r_{0}^{2}-2r_{n}r_{k}\cos(\theta_{n}-\theta_{k})\right)
\right].
\end{equation}

This result is the same as the standard one in Cartesian coordinates
\cite{FH} making the substitution $x=r\cos \theta,\,y=r\sin \theta$.This
result is clear from (\ref{eq:ob}).
Thus, we have shown that, under the constrain of mantaining the paths
$\psi_{k}$ that join points in the discretization

\begin{equation}
  \int{\cal D}x{\cal D}y\exp\left(\frac{i}{\hbar}\int\,dt
\left(\frac{1}{2}(\dot{x}^{2}+\dot{y}^{2})\right)\right)
 =\int{\cal D}r{\cal D}\theta J[r] \exp\left(\frac{i}{\hbar}\int\,dt
\left(\frac{1}{2}(\dot{r}^{2}+r^{2}\dot{\theta}^{2})\right)\right),
 \end{equation}

 where the regularization used is understood.

Explicit application of the conservation of micropaths to define a
covariant path integral may be found in \cite{BePa,Pa} for curved
configuration space using the geodesics as micropaths. For phase space,
there is a very clear exposition for point canonical and other
transformations in \cite{AASA}.

Point  transformations can be made in the naive way so long as
 the micropaths between two points in the discretization
 is preserved. One needs not add extra potential terms to the classsical
Lagrangian when changing coordinates, nor search for strange measures in
order to get the correct result---which just means getting the same
result as in the original integral. Different orderings in the
Hamiltonian operator may be related to the different choices  in the
paths that join two points of the discretization, because this is one of
the freedoms in the definition of the path integral. This freedom is
unrelated whatsoever to that of choosing another set of coordinates in
the Lagrangian to represent the same system. Another freedom is to
choose an equivalent Lagrangian for the system.

We would like to thank Jorge Zanelli for teaching us all we know about
path integrals and enlightening discussions during the course of this
work. Also, we would like to thank F. M\'endez for informative
discussions. M.L. thanks Banco de Chile and A.J. Facultad de Ciencias,
Universidad de Chile, for honour scholarships.

\begin{figure}
\caption{In this figure we can see the limiting procedure with which we
approximate the straight line with semicircles.}
\label{fig1}
\end{figure}

\end{document}